\documentclass[aps,prb,twocolumn,showpacs,showkeys,superscriptaddress,floatfix]{revtex4}
\usepackage{amsmath,graphicx}
\def\integ{\int\displaylimits}
\def\ointeg{\oint\displaylimits}
\def\suma{\sum\displaylimits}

\begin{document}
\title{Anticrossing-induced optical excitonic Aharonov-Bohm effect in strained type-I semiconductor nanorings}

\author{M. Tadi\'c}\email{milan.tadic@etf.bg.ac.rs}
\affiliation{Faculty of Electrical Engineering, University of
Belgrade, PO Box 3554, 11120 Belgrade, Serbia}
\author{N.\v{C}ukari\'c}
\affiliation{Faculty of Electrical Engineering, University of
Belgrade, PO Box 3554, 11120 Belgrade, Serbia}
\author{V. Arsoski}
\affiliation{Faculty of Electrical Engineering, University of
Belgrade, PO Box 3554, 11120 Belgrade, Serbia}
\author{F. M. Peeters}\email{francois.peeters@ua.ac.be}
\affiliation{Department of Physics,  University of Antwerp,
Groenenborgerlaan 171, B-2020 Antwerp, Belgium}

\date{\today}
\begin{abstract}
\normalsize{The exciton states in strained (In,Ga)As nanorings embedded in
a GaAs matrix are computed. The strain distribution is extracted
from the continuum mechanical model, and the exact diagonalization
approach is employed to compute the exciton states. Weak oscillations of the ground exciton state energy with the magnetic field normal to the ring are
an expression of the excitonic Aharonov-Bohm effect. Those oscillations arise from anticrossings between the ground and the second exciton state and can be enhanced by increasing the ring width. Simultaneously, the oscillator strength for exciton recombination exhibits oscillations, which are superposed on a linear increase with magnetic field. The obtained results are contrasted with
previous theoretical results for 1D rings, and differences are explained to arise from different confinement potentials for the electron and the hole, and the large diamagnetic shift present in the analyzed type-I rings. Furthermore, our theory agrees qualitatively well with previous photoluminescence measurements on type-II InP/GaAs quantum dots.}
\end{abstract}


\maketitle

\section{Introduction}

Fabrication of nanometer-sized semiconductor rings triggered interest in the excitonic Aharonov-Bohm effect.\cite{garcia1997,lorke2000,chaplik1995,romer2000} Simple theoretical
models predict oscillations of the exciton levels in one-dimensional
(1D) rings when magnetic field through the ring varies.\cite{chaplik1995,romer2000} Further theoretical work showed no oscillations in the ground exciton state of type-I 2D
and 3D nanorings,\cite{hu2001,ulloa2001} or they were found to be extremely small.\cite{grochol2006} An interesting analytically solvable case is a structure composed of two concentric 1D rings, where the electron and the hole are separately confined, but are Coulomb coupled leading to the formation of exciton. It was found that the oscillator strength for recombination of this exciton could vanish in certain ranges of magnetic field.\cite{govorov2002} However, these bright-to-dark transitions are found only for the case of weak interaction, i.e. when radii of the two rings are small.\cite{govorov2002} Actually, in order to find the optical excitonic AB effect, one should polarize the exciton by confining the electron and the hole in spatially separate potentials. This condition is very difficult to fulfill in type-I semiconductor nanorings, where the electron and the hole are confined in the same space. Hence, to the best of our knowledge no unequivocal experimental confirmation of the optical AB effect for neutral exciton in nanorings has been announced to date.

An appealing and elegant way to polarize the exciton are type-II nanodots,
which confine the electron (hole) inside the dot, whereas the hole (electron)
is expelled to the region outside the dot.\cite{ribeiro2004,kuskovsky2007}
Nevertheless, the latter is confined due to the Coulomb interaction with the former. Such confinement establishes favorable conditions for the occurrence of the optical excitonic Aharonov-Bohm effect. However, no bright-to-dark transitions are found and experiments on different systems show some contradictory details. As an example, Ref.~\onlinecite{kuskovsky2007} found that oscillations in the oscillator strength of stacks of ZnTe/ZnSe nanodots
are superposed on a decaying function of magnetic field. On the other hand, the photoluminescence intensity in a single InP/GaAs quantum dot was found to decrease in narrow ranges of magnetic field, which are arranged periodically, and to increase quasi-linearly between these drops.\cite{degani2008}

In a beautiful experiment on strained type-I (In,Ga)As/GaAs rings, Bayer et al. found Aharonov-Bohm oscillations in the ground state of the charged exciton.\cite{bayer2003} However, the ground state of the
neutral exciton exhibits no oscillations, or they were extremely small
to be experimentally verified. Those rings were fabricated by
means of lithography, and had width of the order of 30 nm. Much narrower
rings are fabricated by means of epitaxy in the Stranski-Krastanov mode,\cite{garcia1997} which allow them to self-assemble on lattice mismatched substrates. Recent experiments, using cross-sectional scanning tunneling microscopy (X-STM), found volcano-like shaped self-assembled rings, with lateral width of 7 nm. The rings are formed from quantum dots, by removing the material in the dot center. The process is driven by strain, and a thin layer of nonuniform thickness resides in the ring opening. Therefore, these rings are not fully opened, which leads to a shift in the transition energy between states of different orbital momenta towards larger magnetic field.\cite{kleemans2007,viefers2004}

\begin{figure}[hbtp]
       \includegraphics[width=6cm]{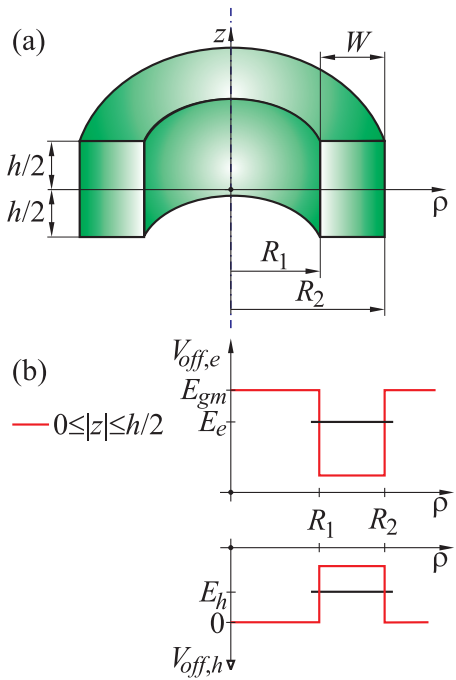}
       \caption{(Color online) (a) The geometry of the ring.
       (b) The confining potential
       due to the band offset as it varies along the $\rho$ axis ($z=0$).
       The electron and hole energy levels, $E_e$ and $E_h$, are
       measured with respect to the top of the valence band in the
       semiconductor matrix.
       \label{fig1}}
\end{figure}

In this paper, the exciton states in an (In,Ga)As nanoring embedded in a GaAs matrix are computed in the presence of a normal magnetic field. The geometry of the ring and its characteristic dimensions are displayed in Fig.~1(a). The analyzed ring is generated by revolving a rectangle of height $h$ and width $W$ about the $z$ axis. The inner radius of the ring is $R_1$ and the outer radius is $R_2$. Energy is measured with respect to the top of the valence band in GaAs, and the energy axis for holes points downwards, as shown in Fig.~1(b). Fig.~1(b) also shows the energy level of the electron $E_e$, the hole energy $E_h$, and energy of the bottom of the conduction band in the GaAs matrix $E_{gm}$. Furthermore, the potentials due to the offsets of the conduction and valence band, $V_{off,e}$ and $V_{off,h}$, respectively, as they vary with $\rho$ for $z=0$ are depicted in Fig.~1(b). The rectangular potential wells shown in Fig.~1(b) are modified by strain, which arises from the lattice mismatch between (In,Ga)As and GaAs. In our approach, the strain distribution is extracted from the continuum mechanical model, and the finite element method (FEM) is employed to discretize the components of the displacement vector on a nonuniform mesh.\cite{tadic2002b} The effective-mass Schr\"odinger equations for the electron and the hole are solved by FEM on the same mesh used to compute the strain distribution. The products of the single-particle wave functions form the basis for the exciton state that are computed within an exact diagonalization scheme. From the exciton wave function, the oscillator strength for exciton recombination is calculated. The exciton energy levels and the oscillator strength are determined as they vary with magnetic field, for a range of the ring width. Our aim is to explore the excitonic Aharonov-Bohm effects in 3D type-I rings, and to investigate their variation with the ring width. A similar model was recently introduced to compute the exciton states in stacks of (In,Ga)As/GaAs rings.\cite{tadic2009}

The paper is organized as follows. Sec. II describes our theoretical approach to compute the electronic structure of the electrons, holes, and excitons. The
numerical results are presented and explained in Sec. III. Our conclusions are given in Sec. IV.

\section{The theoretical model}

Both (In,Ga)As and GaAs are large band-gap semiconductors, thus the
single-band effective mass model can be used  to compute the single-particle states in the conduction band. Tetrahedral deformation of the crystal lattice
due to strain makes the effective potential well for the heavy holes
deeper than the effective potential well for the light holes. Hence,
the ground exciton state is mainly of the heavy-hole origin, and it
justifies use of the single-band effective-mass Hamiltonian
\begin{equation}
    H=T+H_Z+V_{eff},
\label{ham:sing}
\end{equation}
to compute the electron and hole states. Here $T$ denotes the kinetic part
of the Hamiltonian, $H_Z$ is the Zeeman term, and $V_{eff}$ is the effective potential which takes into account both band offset between (In,Ga)As and GaAs and the influence of strain. For our axially symmetric ring, use of cylindrical coordinates $\varphi$, $\rho$ and $z$ is appropriate.

The strain distribution is computed by the 3D continuum mechanical model,
as explained in Ref.~\onlinecite{tadic2002b}. For that purpose, a nonuniform 3D mesh is adopted.\cite{} The computed strain distribution is incorporated
in the effective potential with the assumption of axial symmetry by averaging the strain tensor components over the polar angle.\cite{tadic2002a} Therefore, the effective potentials in the conduction and heavy-hole band depend only on $\rho$ and $z$,
\begin{equation}
    V_{eff}(\rho,z)=V_{off}(\rho,z)+V_{str}(\rho,z),
\end{equation}
where $V_{off}$ denotes the potential due to band offset, and $V_{str}$ is the
strain-dependent effective potential. For the conduction-band electron
\begin{equation}
    V_{str,e}=a_c\left(\varepsilon_{xx}+\varepsilon_{yy}+\varepsilon_{zz}\right),
\end{equation}
and for the heavy hole (hereafter the subscript $h$ is used to denoted the heavy hole)
\begin{equation}
    V_{str,h}=-\left(a_v+\frac{b}{2}\right)
               \left(\varepsilon_{xx}+\varepsilon_{yy}\right)
              -\left(a_v-b\right)\varepsilon_{zz}.
\end{equation}
We compute only the heavy-hole exciton but also test usefulness of the single band approximation for the valence band states by comparing $V_{eff,h}=V_{off,h}+V_{str,h}$ with the effective potential for the light hole
\begin{equation}
    V_{eff,l}=V_{off,h}(\rho,z)-\left(a_v-\frac{b}{2}\right)
               \left(\varepsilon_{xx}+\varepsilon_{yy}\right)
              -\left(a_v+b\right)\varepsilon_{zz}.
\end{equation}
Here, $a_c$, $a_v$, and $b$ denote the deformation potentials, whereas $\varepsilon_{xx}$, $\varepsilon_{yy}$, and $\varepsilon_{zz}$ denote the diagonal components of the strain tensor.

The kinetic part of the Hamiltonian is written in the symmetric gauge
\begin{equation}
    \begin{split}
    T&=-\frac{\hbar^2}{2m_\parallel}\frac{\partial^2}{\partial\rho^2}
    -\frac{\hbar^2}{2m_\parallel}\frac{1}{\rho}\frac{\partial}{\partial\rho}
    -\frac{\hbar^2}{2m_\parallel}\frac{1}{\rho^2}\frac{\partial^2}{\partial\varphi^2}\\
    &-\frac{\hbar^2}{2}\frac{\partial}{\partial\rho}
    \left(\frac{1}{m_\parallel}\right)\frac{\partial}{\partial\rho}\\
    &-\frac{\hbar^2}{2m_z}\frac{\partial^2}{\partial z^2}
    -\frac{\hbar^2}{2}\frac{\partial}{\partial z}
    \left(\frac{1}{m_z}\right)\frac{\partial}{\partial z}\\
    &+\frac{\hbar^2}{2m_\parallel}\frac{1}{4l_c^4}\rho^2
    -\frac{\hbar^2}{2m_\parallel}\frac{i}{l_c^2}\frac{\partial}{\partial\varphi},
    \end{split}
\label{t}
\end{equation}
where $l_c=(\hbar/eB)^{1/2}$ denotes the magnetic length for the magnetic field $B$. $m_\parallel$ and $m_z$ in Eq.~(\ref{t}) are the effective masses in the $xy$ plane and along the $z$ direction, respectively. $m_\parallel=m_z$ for the electron in the conduction band, while for the heavy hole $m_\parallel$ and $m_z$ are extracted from the diagonal approximation of the multiband Luttinger-Kohn model.\cite{tadic2002a,vurgaftman2001}

The Zeeman term has the form
\begin{equation}
    H_{Z,e}=\pm g_{eff}\mu_BB/2,
\label{hze}
\end{equation}
\begin{equation}
    H_{Z,h}=\mp 3\kappa\mu_BB,
\label{hzh}
\end{equation}
for the electrons and heavy holes, respectively. Here, $\mu_B$ denotes the Bohr magneton, $g_{eff}$ denotes the effective Land\'e $g$-factor, and $\kappa$ is the Luttinger parameter describing the Zeeman splitting of the hole states, and the upper (lower) sign in Eqs.~(\ref{hze}) and (\ref{hzh}) refer to the spin-up (spin-down) electron states.

In order to solve the single-band effective-mass Schr\"odinger equation,
$H\Psi=E\Psi$, the Galerkin form of the finite element method is employed.
Our calculations rely on
\begin{equation}
    \integ_V\left[u{\rm div}{\bf A}+({\bf A}\cdot{\rm grad})u\right]dV
    =\ointeg_S u{\bf A}\cdot d{\bf S},
\label{green}
\end{equation}
where $V$ denotes the solution domain, $u$ is an arbitrary scalar function, ${\bf A}$ is an arbitrary vector-valued function, and $S$ is the boundary of $V$. When applied to the Hamiltonian (\ref{ham:sing}), Eq.~(\ref{green}) gives
\begin{equation}
    \begin{split}
    &\integ_V u (\tilde{T}+\tilde{H}_Z+\tilde{V}_{eff})\Psi d{\bf r}\\
    &=\integ_V \frac{m_0}{m_\parallel}\left(
    \frac{1}{\rho^2}\frac{\partial u}{\partial\varphi}
    \frac{\partial \Psi}{\partial\varphi}+
    \frac{\partial u}{\partial \rho}\frac{\partial \Psi}{\partial\rho}\right)d{\bf r}\\
    &+\integ_V\frac{m_0}{m_z}\left(\frac{\partial u}{\partial z}\frac{\partial \Psi}{\partial z}\right)d{\bf r}
    +\integ_V u \tilde{H}_B^\prime\Psi d{\bf r}+\integ_V u\tilde{V}\Psi d{\bf r}.
    \end{split}
\end{equation}
Here, $\tilde{T}=T/(\hbar^2/2m_0)$, $\tilde{H}_Z=H_Z/(\hbar^2/2m_0)$, and
$\tilde{V}_{eff}=V_{eff}/(\hbar^2/2m_0)$.

The single-particle Hamiltonian is axially symmetric, and therefore the
projection of the orbital quantum number on the $z$ axis $L_z=l\hbar$ is
a good quantum number for both the electron and hole states. For a given $l$, the single-particle states are denoted by the principal
quantum number $n$ and the parity $\sigma$, i.e. $nl^\sigma$. They could
additionally be labeled by spin, but we are interested in the (electron) spin-up states in the two bands whose eigenenergies are lower than the energies of the spin-down states. The wave function of the $nl^\sigma$ state is written as $\Psi_{ln}^\sigma$, and the eigenenergy as $E_{ln}^\sigma$. The energies of the ground electron
and hole states, whose orbital momentum varies with $B$, are denoted by
the symbols $E_e^{(1)}$ and $E_h^{(1)}$.

Because of axial symmetry, the single particle wave function of the $nl^\sigma$ state can be written as
\begin{equation}
    \Psi_{ln}^\sigma(\varphi,\rho,z)=\frac{1}{\sqrt{2\pi}}e^{il\varphi}\psi_{ln}^\sigma(\rho,z),
\label{psi:psi}
\end{equation}
where $\psi_{l,n}(\rho,z)$ is expanded in the first-order shape functions
\begin{equation}
    \psi_{ln}^\sigma(\rho,z)=\sum_{jk}c_{jk}f_j(\rho)f_k(z),
\label{psi:f}
\end{equation}
which are labeled by the mesh points, $j$ and $k$. On the master element $[-1,1]$, the first-order shape function has the form
\begin{equation}
    f(\xi)=\begin{cases}
           (1+\xi)/2,&-1\leq\xi\leq 0\\
           (1-\xi)/2,&0\leq\xi\leq +1
           \end{cases}.
\end{equation}
Eq.~(\ref{psi:f}) leads to the generalized eigenvalue problem
\begin{equation}
    {\bf H}{\bf c}=E{\bf S}{\bf c}.
\end{equation}
The matrix elements of ${\bf H}$ and ${\bf S}$ are given by $H_{ij}=\langle f_i\vert H\vert f_j\rangle$ and $S_{ij}=\langle f_i\vert f_j\rangle$.

The exciton states are extracted from the equation
\begin{equation}
    H_x\Psi_x=(H_e+H_h+V_{C})\Psi_x=E_x\Psi_x,
\label{psi:x}
\end{equation}
where $H_x$ denotes the exciton Hamiltonian, $E_x$ is the
exciton eigenenergy, $\Psi_x=\Psi_x({\bf r}_e,{\bf r}_h)$ is
the exciton wave function, $H_e$ and $H_h$ are the electron and the hole
Hamiltonian, respectively, and $V_C$ is the Coulomb potential,
\begin{equation}
    V_C({\bf r}_e,{\bf r}_h)=-\frac{e^2}{4\pi\epsilon_s\epsilon_0\sqrt{\rho_{x}^2+(z_e-z_h)^2}}.
\end{equation}
Here $\epsilon_0$ is the vacuum permittivity, $\epsilon_s$ is the relative
permittivity of the material inside the ring, $z_e$ and $z_h$ are the
values of the $z$ coordinate of the electron and the hole,
whereas $\rho_x$ denotes the projection of the distance between
the electron and the hole on the $xy$ plane
\begin{equation}
    \rho_x=\left[\rho_e^2+\rho_h^2-2\rho_e\rho_h\cos(\varphi_e-\varphi_h)\right]^{1/2}.
\label{rhox}
\end{equation}
From now on, $\rho_x$ will be referred as the in-plane distance between the electron and the hole.

Let us briefly examine which quantum numbers label the exciton states. The in-plane distance $\rho_x$ does not depend on the polar angles of the electron and the hole, $\varphi_e$ and $\varphi_h$, separately,  but on the difference $\varphi_e-\varphi_h$. It implies that rotation of the exciton as a whole over the $z$ axis by an arbitrary angle does not affect the Coulomb interaction, and the orbital momentum $L=l_e+l_h$ is a good quantum number of the exciton. Furthermore, $H_x$ possesses the inversion symmetry with respect to simultaneous reversal of the $z_e$ and $z_h$ coordinates.
Therefore, the exciton parity $\sigma_{x}$ is a good quantum number.
The even and odd exciton states, $\sigma_x=+$ and $\sigma_x=-$, respectively, are composed of the electron and hole states of equal and opposite parity, respectively. For the given exciton and electron parities, $\sigma_x$ and $\sigma_e$, respectively, the hole parity $\sigma_h=\sigma_h(\sigma_x,\sigma_e)$ has the following values
\begin{equation}
    \begin{split}
    &\sigma_h(+,+)=+,\mskip 12mu
    \sigma_h(+,-)=-,\\
    &\sigma_h(-,+)=-,\mskip 12mu
    \sigma_h(-,-)=+.
    \end{split}
\end{equation}

At zero magnetic field the exciton states are arranged in spin quartets
$[\uparrow\uparrow,\uparrow\downarrow,\downarrow\uparrow,\downarrow\downarrow]$,
where the first arrow indicates the spin of the conduction-band state and the second arrow denotes the spin of the valence-band state. In magnetic field, the Zeeman terms in Eq.~(\ref{psi:x}) split the spin quartets so that the $\uparrow\uparrow$ excitons have the lowest energies among their counterparts. Furthermore, energies of the odd exciton states are higher by a few tens of meV from those of the even exciton states. Therefore, only even exciton states of the spin-up electron and spin-up hole are presented and discussed in Sec. III.
The exciton states are  denoted by $nL^{\sigma_{x}}$, and the exciton
eigenenergies by $E_{xnL}^{\sigma_x}$, where $n$ denotes the principal quantum number. For the energy of the ground exciton state we use the abbreviated symbol $E_x^{(1)}$.

The Schr\"odinger equation for the exciton reads
\begin{equation}
    \Psi_x=\suma_{\sigma_e}\sum_{l_e}\sum_{n_e}\sum_{n_h}c_{l_e,n_e,n_h}^{\sigma_e}
    \Psi_{l_e,n_e}^{\sigma_e}({\bf r}_e)\Psi_{l_h,n_h}^{\sigma_h}({\bf r}_h),
\end{equation}
where $l_h=L-l_e$, and $\sigma_h=\sigma_h(\sigma_x,\sigma_e)$. Our exact diagonalization approach extracts the exciton states from the secular equation
\begin{equation}
    \begin{split}
    &(E_{gm}+E_{l_e,n_e}^{\sigma_e}+E_{l_h,n_h}^{\sigma_h}-E_{x})
    \delta_{\sigma_e^\prime,\sigma_e}\delta_{l_e^\prime,l_e}\delta_{n_e^\prime,n_e}
    \delta_{n_h^\prime,n_h}\\
    &+\sum_{\sigma_e}\sum_{l_e}\sum_{n_e}\sum_{n_h}
    \langle\Psi_{l_e^\prime,n_e^\prime}^{\sigma_e^\prime}\Psi_{l_h^\prime,n_h^\prime}^{\sigma_h^\prime}\vert
    V_{C}\vert\Psi_{l_e,n_e}^{\sigma_e}\Psi_{l_h,n_h}^{\sigma_h}\rangle=0,
    \end{split}
\label{exc:sec}
\end{equation}
where $\delta$ denotes the Kronecker delta and $E_{gm}$ is the energy gap in the GaAs matrix. A straightforward derivation gives
\begin{equation}
    \begin{split}
    &\langle\Psi_{l_e^\prime,n_e^\prime}^{\sigma_e^\prime}
            \Psi_{l_h^\prime,n_h^\prime}^{\sigma_h^\prime}\vert
    V_{C}\vert\Psi_{l_e,n_e}^{\sigma_e}\Psi_{l_h,n_h}^{\sigma_h}\rangle=
    \frac{1}{4\pi^2}\frac{e^2}{\epsilon\epsilon_0}
    \delta_{l_{e}^\prime+l_{h}^\prime,l_{e}+l_{h}}\\
    &\times\integ_0^\infty dk_\parallel k_\parallel
    \integ_{-\infty}^{+\infty} dk_z
    \frac{1}{k_\parallel^2+k_z^2}
    {\cal F}_{e}(k_\parallel,-k_z){\cal F}_{h}(k_\parallel,k_z).
    \end{split}
\label{exc:coul}
\end{equation}
where $k_\parallel$ and $k_z$ denote the in-plane and the $z$
component of the wave vector in Fourier space, respectively.
${\cal F}_{e}$ is the two-dimensional transform given by:
\begin{equation}
    {\cal F}_e(k_\parallel,k_z)=\integ_0^\infty\rho d\rho
        \integ_{-\infty}^{+\infty}
        dz\psi_{l_e^\prime,n_e^\prime}^{\sigma_e^\prime\ast}\psi_{l_e,n_e}^{\sigma_e} J_{\vert
        l_e^\prime-l_e\vert}(k_\parallel\rho)e^{ik_zz},
\label{exc:matel}
\end{equation}
with $J_l(x)$ the Bessel function of the first kind.

As a figure of merit of the exciton, we compute the average exciton
in-plane radius $R_\parallel$, whose square is given by
\begin{equation}
    R_\parallel^2=\langle \rho_x^2\rangle=
    \integ_{V_e}d{\bf r}_e\integ_{V_h}
    \rho_x^2({\bf r}_e,{\bf r}_h)
    \vert\Psi_x({\bf r}_e,{\bf r}_h)\vert^2d{\bf r}_h.
\end{equation}
where $\rho_x({\bf r}_e,{\bf r}_h)$ is given in Eq.~(\ref{rhox}). Replacing the single particle wave function with the form in Eq.~(\ref{psi:psi}), and taking into account parity, results in
\begin{equation}
    \begin{split}
    R_\parallel^2&=\suma_{\sigma_e}\suma_{l_e}\suma_{n_e^\prime,n_e}\suma_{n_h^\prime,n_h}
    c_{l_e,n_e^\prime,n_h^\prime}^{\sigma_e\ast}c_{l_e,n_e,n_h}^{\sigma_e}\\
    &\times\left[
    \langle l_e,n_e^\prime\vert \rho_e^2\vert l_e,n_e\rangle
    \delta_{l_h^\prime,l_h}+\delta_{l_e^\prime,l_e}
    \langle l_h,n_h^\prime\vert \rho_h^2\vert l_h,n_h\rangle\right]\\
    &-c_{l_e-1,n_e^\prime,n_h^\prime}^{\sigma_e\ast}c_{l_e,n_e,n_h}^{\sigma_e}\\
    &\times\langle l_e-1,n_e^\prime\vert \rho_e\vert l_e,n_e\rangle
    \langle l_h+1,n_h^\prime\vert \rho_h\vert l_h,n_h\rangle\\
    &-c_{l_e+1,n_e^\prime,n_h^\prime}^{\sigma_e\ast}c_{l_e,n_e,n_h}^{\sigma_e}\\
    &\times\langle l_e+1,n_e^\prime\vert \rho_e\vert l_e,n_e\rangle
    \langle l_h-1,n_h^\prime\vert \rho_h\vert l_h,n_h\rangle.
    \end{split}
\label{rpar2}
\end{equation}
where $l_h=L-l_e$. The matrix elements in Eq.~(\ref{rpar2}),
\begin{equation}
    \langle l^\prime,n^\prime\vert \rho^k\vert l,n\rangle=\integ_{-H/2}^{H/2} dz
    \integ_{0}^R\rho^k\psi_{l^\prime n^\prime}^\ast(\rho,z)
    \psi_{ln}(\rho,z)\rho d\rho,
\end{equation}
are computed numerically for $k=0,1,2$ on the solution domain of radius $R$ and height $H$. In addition to $R_\parallel$, we compute the binding
energy of the exciton
\begin{equation}
    E_b=E_{gm}+E_{e}^{(1)}+E_{h}^{(1)}-E_{x}^{(1)}.
\end{equation}

The oscillator strength for exciton recombination is given by
\begin{equation}
    f_{x}=\frac{2}{m_0E_{xi}}\vert\langle u_{c0}\vert{\boldsymbol \varepsilon}\cdot {\bf p}\vert u_{v0}\rangle\vert^2
    \vert M\vert^2.
\label{fx:gen}
\end{equation}
Here, ${\boldsymbol\varepsilon}$ denotes the unit vector of polarization
of outcoming light, $u_{c0}$ and $u_{v0}$ are the periodic parts of the
Bloch functions of the electron in the conduction and valence band, respectively, ${\bf p}$ is the electron momentum, $E_x$ is the exciton energy, $m_0$ is the free-electron mass, and $M$ denotes the transition matrix element between the envelope functions\cite{efros1996}
\begin{equation}
    M=\integ_{V_e,V_h}
    \delta({\bf r}_e-{\bf r}_h)\Psi_{x}({\bf r}_e,{\bf r}_h)d{\bf r}_e d{\bf r}_h.
\label{exc:m1}
\end{equation}
For equal spins of the electron and the hole, and even parity of the exciton,
only $L=l_e+l_h=0$ exciton states are bright, therefore
\begin{equation}
    M=\suma_{\sigma_e}\suma_{l_e}
    \suma_{n_e}\suma_{n_h}  c_{l_e,n_e,n_h}^{\sigma_e}\langle l_e,n_e\vert -l_e,n_h\rangle.
\label{exc:m2}
\end{equation}
We assume light polarized along the $x$ direction, for which
the matrix element squared between the zone center states is given by\cite{davies1997,climente2003}
\begin{equation}
    \vert\langle u_{c0}\vert p_x\vert u_{v0}\rangle\vert^2=\frac{m_0^2P^2}{2\hbar^2},
\label{exc:p}
\end{equation}
where $P$ denotes the Kane interband matrix element. When Eqs.~(\ref{exc:m1})$-$(\ref{exc:p}) are inserted in Eq.~(\ref{fx:gen}), the expression for $f_{x}$ of a bright exciton state follows
\begin{equation}
    f_{x}=\frac{1}{2}\frac{E_P}{E_{xi}}\left\vert\suma_{\sigma_e}\suma_{l_e}\suma_{n_e}\suma_{n_h}  c_{l_e,n_e,n_h}^{\sigma_e}
    \langle l_e,n_e\vert -l_e,n_h\rangle\right\vert^2,
\end{equation}
where $E_P=2m_0P^2/\hbar^2$. At finite temperature, the dark states are occupied with a finite probability. One defines the (dimensionless) photoluminescence intensity,\cite{degani2008}
\begin{equation}
    I_{PL}=\frac{\suma_i f_{xi}\exp{\left(-E_{xi}/k_BT\right)}}
    {\suma_i\exp{\left(-E_{xi}/k_BT\right)}},
\end{equation}
which takes into account that the exciton states, labeled by a single index $i$, are populated according to
Boltzmann statistics.

\section{Numerical results and discussion}

We compute the exciton states in the (In,Ga)As nanoring embedded in the GaAs matrix. Such rings have been recently fabricated and analyzed.\cite{kleemans2007} X-STM analysis revealed they have nearly circular cross section with inner and outer radii $R_1=8$ nm and $R_2=15$ nm and height  $h=4$ nm. Similarly, in our calculations $R_1$ equals 8 nm, while the ring width $W$ is varied in the range from 2 to 22 nm. We assumed that the mole fraction of InAs in the ring is $x=0.55$.\cite{kleemans2007} The parameters of the band structure and elastic constants of (In,Ga)As and GaAs are all taken
from Ref.~\onlinecite{vurgaftman2001}. The band offset is such that 83$\%$ of the band-gap difference is realized in the conduction band.\cite{stier1999} The nonuniform mesh in the finite-element calculation of the single-particle states is constructed from 129 points along both the $\rho$ and $z$ direction. The expansion domain is $H=200$ nm high and its radius is $R=120$ nm. $g_{eff}$ and $\kappa$ are taken to be position independent and equal to the values in GaAs, -0.44 and 1.2, respectively. Our choice is supported by experiments which indicated that the energy level splitting in (In,Ga)As dots is much smaller than what is found in bulk (In,Ga)As, and that they are closer to the values in the GaAs matrix.\cite{bayer1999} The basis for the exciton states is constructed from 6 even and 2 odd single-particle states for each $l$ ($l_e$ or $l_h$) in the range from -7 to +7. We assumed a temperature $T=1$ K.

\begin{figure}[hbtp]
       \includegraphics[width=8cm]{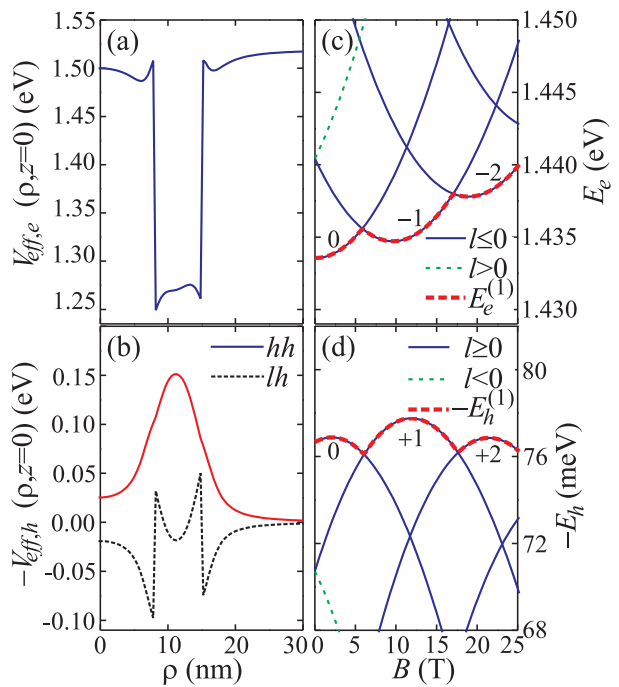}
       \caption{(Color online) (a) The effective potential for the electron
       along the radial axis for $z=0$. (b) The same for the heavy-hole
       (solid line) and the light-hole (dashed line). (c) The electron energy levels as they vary with magnetic field. (d) Negative of the heavy-hole eigenenergies as function of magnetic field. The ground states $E_{e}^{(1)}$ and $E_{h}^{(1)}$ oscillate with magnetic field, and the orbital momentum changes, as indicated by numbers adjacent to the curves. The ring has a width 7 nm and a height 4 nm.
       \label{fig2}}
\end{figure}

The effective potentials in the conduction and valence bands of $W=7$ nm
wide nanoring along the $\rho$ direction are shown in Figs. 2(a) and 2(b),
respectively. The effective potential well for the electron is deeper
than the effective potential well for the heavy hole.  On the other
hand, the heavy hole is confined in a wide effective potential well,
which is much deeper than the effective potential well for the light hole.
It turns out that the energy levels of the light hole are pushed by strain further from the heavy-hole energy levels towards the continuum. Consequently, strain reduces mixing between the light holes and the heavy holes,\cite{tadic2002a} which supports the use of the diagonal approximation of the Luttinger-Kohn model when computing the hole states. Variations of the electron and hole states with magnetic field are shown in Figs.~2(c) and 2(d). Both the electron and hole energy levels show orbital momentum transitions, which take place at almost the same magnetic field values. Therefore, to a great certainty we may infer that the orbital momentum of the ground exciton state is $L=l_e+l_h=0$ irrespective of the magnetic field value.

\begin{figure}[hbtp]
       \includegraphics[width=8cm]{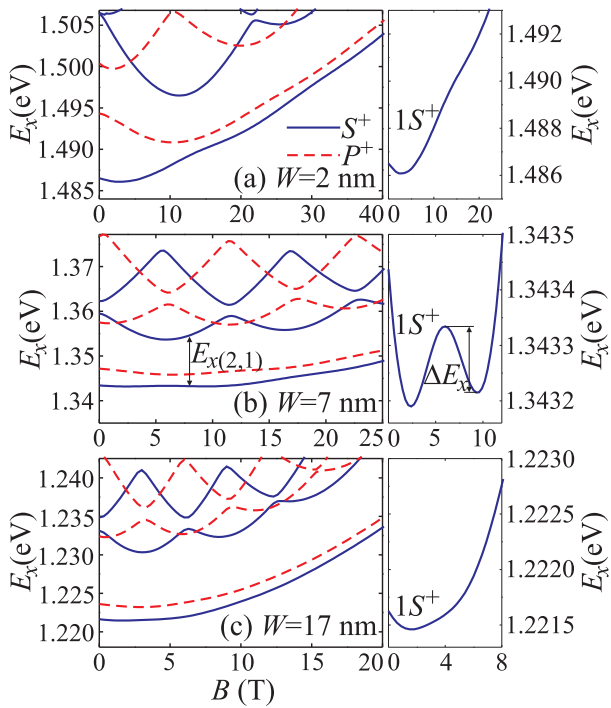}
       \caption{(Color online) The lowest-energy $L=0$ (solid lines)
       and $L=-1$ states (dashed lines) in: (a) the $W=2$, (b) the $W=7$, and (c) the $W=17$ nm wide ring. Two minima of the ground exciton states are found in only the 7 nm wide ring. Right panel displays detailed
       views of the ground exciton energy variations at
       low magnetic fields.
       \label{fig3}}
\end{figure}

The ground exciton state $E_x^{(1)}$ is indeed $E_{x01}^+$, as Figs.~3(a), (b), and (c) show for the $W=2$, $W=7$, and $W=17$ nm wide ring, respectively.
Details of $E_{x}^{(1)}$ as function of $B$ for the three cases are shown in the right panel of Fig.~3, and indicate that the width of the ring affects the energy variation of the ground exciton state with magnetic field. While the cases of narrow and wide rings, Figs.~3(a) and (c), do not clearly demonstrate oscillations of $E_x^{(1)}$, they become evident in the $W=7$ nm wide ring (see Fig.~3(b)). The first minimum of $E_x^{(1)}$ as function of $B$ in Figs.~3(a)$-$(c) arises from anticrossing with the $E_{x02}^+$ state, and is affected by the Zeeman splitting at low magnetic field, when the diamagnetic shift is not large. Out of the three curves in the right panel of Fig.~3, only $E_x^{(1)}$ dependence on $B$ in the $7$ nm wide ring exhibits two minima. By comparing Figs.~2 and 3, we see that the anticrossings of the exciton states take place close to the orbital momentum transitions of the single particle states.

The energy difference between $E_{x02}^+$ and $E_{x01}^+$, i.e. $E_{x(2,1)}=E_{x02}^+-E_{x01}^+$ is explicitly indicated in Fig.~3(b).
The other parameter is $\Delta E_x$, the difference between the first
maximum and the second minimum of the $E_{x01}^+(B)$ curve, which
is indicated in the right panel of Fig.~3(b). If $\Delta E_x$ could be defined
(only for the $W=7$ nm wide ring in Fig.~3), oscillations in the energy of
the $1S^+$ exciton state are clearly visible. As Fig.~3(a) shows, the Aharonov-Bohm oscillations in the narrow 2 nm wide ring are suppressed by large diamagnetic shift. On the other hand, the confinement of the single-particle states in the 17 nm wide ring becomes strong and the Coulomb interaction weak, therefore no large oscillations are observed in Fig.~3(c). For the intermediate case, shown in Fig.~3(b), the oscillations do not suffer from either diamagnetic shift or strong confinement, which establishes favorable conditions for the appearance of the second minimum in the $E_x^{(1)}$ dependance on $B$. Moreover, irrespective of the ring width, quite large oscillations, with amplitude of the order of 10 meV, exist in the higher exciton energy levels as they depend on $B$.

\begin{figure}[hbtp]
       \includegraphics[width=5.5cm]{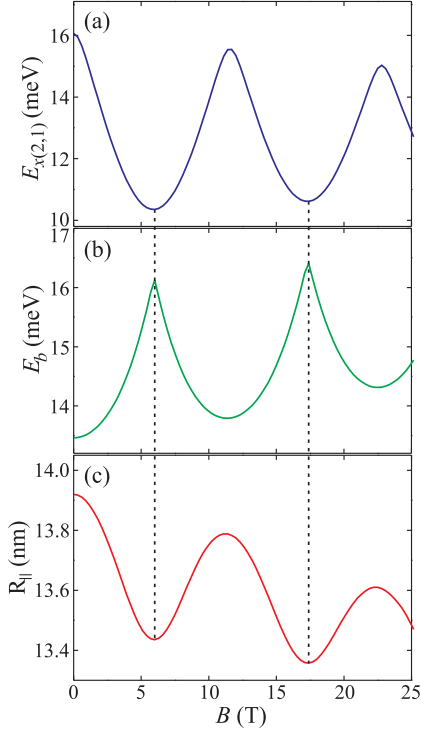}
       \caption{(Color online) (a) The difference between two lowest
       energy $L=0$ exciton states, $E_{x(2,1)}$, (b) the binding energy
       of the ground exciton state $E_b$, and (c) the average in-plane radius
       $R_\parallel$ as they vary with $B$ in the 7 nm wide ring.
       The minima in (a) and (c) correspond to maxima in (b), which
       is indicated by the dashed lines.
       \label{fig4}}
\end{figure}

Fig.~4(a) shows variation of $E_{x(2,1)}$ with $B$ in the $W=7$ nm wide ring.
It is oscillatory with the minima corresponding to anticrossings between
the $E_{x02}^+$ and $E_{x01}^+$ states. The ratio of magnetic field values where these minima take place is close to 1:3:5:..., which is the exact order of the
single-particle orbital momentum transitions in 1D rings.\cite{viefers2004}
The radius of the equivalent 1D ring estimated from the magnetic field interval between two minima $\Delta B$ in Fig.~4(a) is $R_{1D}=\left(h/e\pi\Delta B\right)^{1/2}$=10.7 nm, which is close to the average of the inner and outer radius of the ring, $(R_1+R_2)/2=11.5$ nm. Each minimum of $E_{x(2,1)}$ corresponds to a maximum of the binding energy $E_b$, as shown in Fig.~4(b). Furthermore, due to non-smooth variation of $E_{h}^{(1)}$ and $E_{e}^{(1)}$ (see Fig. 2), $E_b$ exhibits spikes at anticrossings. When $E_b$ is large, the Coulomb interaction is large, and the electron and the hole are bound close to  each other, as demonstrated in Fig.~4(c). The oscillations of $R_\parallel$ are clearly observed in Fig.~4(c), although the amplitude of these oscillations is not large.

\begin{figure}[hbtp]
       \includegraphics[width=5.5cm]{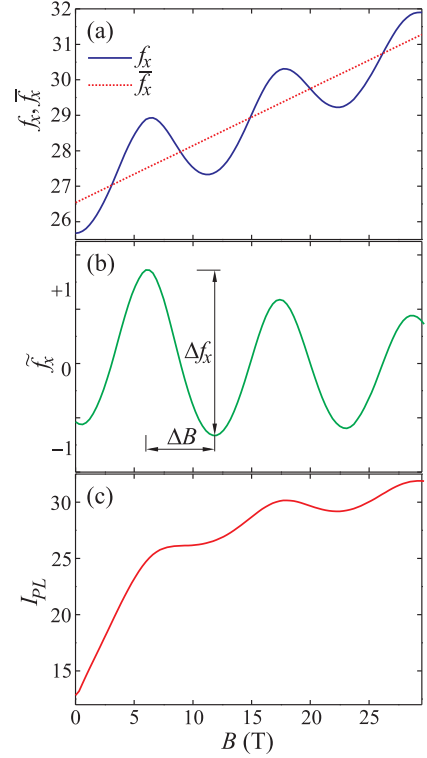}
       \caption{(Color online) (a) The oscillator strength for recombination of the ground exciton state $f_x$ (solid line) and the linear fitting curve $\overline{f}_x$ (dashed line) as function of magnetic field. (b) Variation of the residue $\tilde{f}_x=f_x-\overline{f}_x$ with magnetic field. (c) The photoluminescence intensity $I_{PL}$ exhibits weak oscillations when magnetic field varies. The ring width equals $W=7$ nm and the height is $h=4$ nm.
       \label{fig5}}
\end{figure}

The Aharonov-Bohm oscillations give rise to oscillations in $f_x$, which are shown for the ground exciton energy level in Fig. 5(a). Increasing the magnetic field leads to a decrease of $R_\parallel$, which in turn leads to an increase of $f_{x}$. When $R_\parallel(B)$ has a minimum, $f_{x}(B)$ achieves a maximum. Variation of $f_{x}$ with $B$ shown in Fig.~5(a) seems to have the form
$f_{x}=\overline{f}_x+\tilde{f}_x$, where $\overline{f}_x=aB+b$ is a linear function of $B$, and $\tilde{f}$ is the oscillatory residue. $\overline{f}_{x}(B)$ is displayed by the dashed straight line in Fig.~5(a), while $\tilde{f}_x=f_{x}-\overline{f}$ as function of $B$ is shown in Fig.~5(b). The amplitude of oscillations is defined as the difference between the first maximum and the first minimum of $f_x$, which is denoted by $\Delta f_x$ in Fig.~5(b). Furthermore, $\Delta B$ in Fig.~5(b) denotes the interval of magnetic field between the first minimum and the first maximum (see Figs.~5(a) and 5(b)). Both $\Delta f_x$ and $\Delta B$ are used to compute the relative amplitude of the first oscillation
\begin{equation}
    \delta f_x=\Delta f_x/\langle f_x\rangle,
\label{d:fx}
\end{equation}
where $\langle f_x\rangle$ is the average value of $f_x$ in the interval $\Delta B$. Due to the population of higher exciton states, the oscillations of $I_{PL}$ are considerably smeared out, even at temperature as low as 1 K (see Fig.~5(c)). Such small oscillations might be very difficult to observe experimentally, but they resemble the Aharonov-Bohm oscillations of the photoluminescence intensity measured in type-II InP/GaAs quantum dots.\cite{degani2008}

\begin{figure}[hbtp]
       \includegraphics[width=7cm]{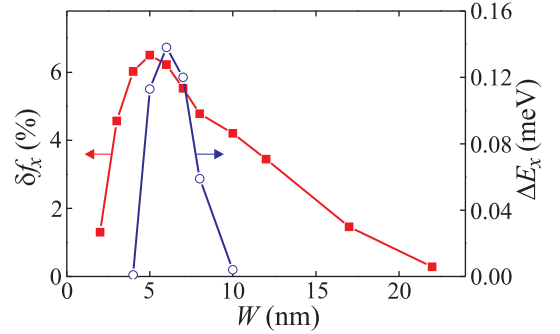}
       \caption{(Color online) Variations of $\Delta E_x$ (blue circles) and
       $\delta f_x$ (red squares) with the ring width.
       \label{fig6}}
\end{figure}

Figs.~3(a)-(c) illustrate that the magnitude of the Aharonov-Bohm oscillations of the ground exciton energy level depends on the width of the analyzed type-I ring. As a matter of fact, Fig.~6 shows that $\Delta E_x$ and $\Delta f_x$ of the ground exciton state are subject to changes when $W$ varies. The maximum of the $\Delta E_x(W)$ curve equals 0.14 meV, and is located at $W=6$ nm, while $\delta f_x$ exhibits a maximum of 6.5$\%$ which is located at $W=5$ nm. $\delta f_x$ is a well defined property of the ground exciton state in the whole explored range of $W$, from 2 to 22 nm. On the other hand, $\Delta E_x>0$ and $E_x^{(1)}$ vs $B$ dependence exhibits a second minimum only if the ring width is in the range from 4 to 10 nm. Different domains of $\delta f_x$ and $\Delta E_x$ imply that the optical excitonic Aharonov-Bohm effect is present in type-I (In,Ga)As semiconductor nanorings, even though oscillations of the ground exciton level are not clearly visible. Previous analysis of concentric 1D rings showed that oscillations of the oscillator strength arise due to periodical bright to dark transitions of the exciton states, therefore $\delta f_x$ is much larger in concentric 1D rings. In the analyzed rings, the electron and the hole are localized in the same space, thus the exciton is only weakly polarized, and no bright-to-dark transitions are found. 3D rings, therefore, offer different physics of the optical excitonic Aharonov-Bohm effect than concentric 1D rings.\cite{govorov2002}

\section{Conclusion}

We show theoretically that both the excitonic and optical excitonic Aharonov-Bohm effects are present in strained type-I (In,Ga)As/GaAs nanorings. The Aharonov-Bohm oscillations of the exciton ground state arise from anticrossings between the exciton energy levels, which change the exciton radius, and therefore bring about oscillations in the oscillator strength for exciton recombination. For rings of experimental inner radius and height, the amplitude of these oscillations is found to depend on the ring width. Our calculations show that a large diamagnetic shift suppresses the oscillations of both the energy levels and the oscillator strength when the ring width is of the order of 2 nm. Similarly, oscillations in wide rings (whose width is of the order of 20 nm) become negligible. The maximum amplitude of oscillations of about 0.14 meV in the ground exciton energy level is realized for the $6$ nm wide ring. The oscillator strength for exciton recombination exhibits oscillations around a quasi-linear dependence on the magnetic field, which is qualitatively similar to the dependence observed in Ref.~\onlinecite{degani2008}. These oscillations are, however, only a few percent of the average value of the oscillator strength. Our calculations indicate that with a proper design of the type-I nanoring, e.g. varying its dimensions, one can realize an enhancement of both the excitonic and
optical excitonic Aharonov-Bohm effects.

\section*{Acknowledgment}

This work was supported by the Ministry of Science of Serbia,
the Flemish Science Foundation (FWO-Vl), the EU NoE: SANDiE, and
the Belgian Science Policy (IAP). The calculations were performed
on the CalcUA and Seastar computer clusters of the University of Antwerp.


\begin{thebibliography}{00}
\bibitem{garcia1997}J. M. Garcia, G. Medeiros-Ribeiro, K. Schmidt,
T. Ngo, J. L. Feng, A. Lorke, J. Kotthaus, and P. M. Petroff,
Appl. Phys. Lett. {\bf 71}, 2014 (1997).
\bibitem{lorke2000}A. Lorke, R. Johannes Luyken, A. O. Govorov, J.
P. Kotthaus, J. M. Garcia, and P. M. Petroff, Phys. Rev. Lett.
{\bf 84}, 2223 (2000).
\bibitem{chaplik1995}A. V. Chaplik, JETP Lett. 62, 900 (1995).
\bibitem{romer2000}R. A. R\"omer and M. R. Raikh, Phys. Rev. B
{\bf 62}, 7045 (2000).
\bibitem{hu2001}H. Hu, J. L. Zhu, D. J. Li, and J. J. Xiong, Phys.
Rev. B {\bf 63}, 195307 (2001).
\bibitem{ulloa2001}J. Song and S. E. Ulloa, Phys. Rev. B {\bf 63},
125302 (2001).
\bibitem{grochol2006}M. Grochol, F. Grosse, and R. Zimmermann,
Phys. Rev. B {\bf 74}, 115416 (2006).
\bibitem{govorov2002}A. O. Govorov, S. E. Ulloa, K. Karrai, and R.
J. Warburton, Phys. Rev. B {\bf 66}, 081309 (2002).
\bibitem{ribeiro2004}E. Ribeiro, A. O. Govorov, W. Carvalho, Jr.,
and G. Medeiros-Ribeiro, Phys. Rev. Lett. {\bf 92}, 126402 (2004).
\bibitem{kuskovsky2007}I. L. Kuskovsky, W. MacDonald, A. O. Govorov, L. Mourokh,
X. Wei, M. C. Tamargo, M. Tadi\'c, and F. M. Peeters, Phys. Rev. B
{\bf 76}, 035342 (2007).
\bibitem{degani2008}M. H. Degani, M. Z. Maialle, G. Medeiros-Ribeiro,
and Evaldo Ribeiro, Phys. Rev. B {\bf 78}, 075322 (2008).
\bibitem{bayer2003}M. Bayer, M. Korkusinski, P. Hawrylak, T. Gutbrod,
M. Michel, and A. Forchel, Phys. Rev. Lett. {\bf 90}, 186801 (2003).
\bibitem{kleemans2007}N. A. J. M. Kleemans, I. M. A. Bominaar-Silkens,
V. M. Fomin, V. N. Gladilin, D. Granados, A. G. Taboada, J. M.
Garc\'ia, P. Offermans, U. Zeitler, P. C. M. Christianen, C. J.
Maan, J. T. Devreese, and P. M. Koenraad, Phys. Rev. Lett. {\bf 99}
146808 (2007).
\bibitem{viefers2004} S. Viefers, P. Koskinen, P. Singha Deo, and M. Manninen,
Physica E {\bf 21}, 1 (2004).
\bibitem{tadic2002b}M. Tadi\'c, F. M. Peeters, K. L.
Janssens, M. Korkusi\'nski, and P. Hawrylak, J. Appl. Phys. {\bf
92}, 5819 (2002).
\bibitem{tadic2009}M. Tadi\'c and F. M. Peeters, Phys. Rev. B {\bf 79},
153305 (2009).
\bibitem{tadic2002a}M. Tadi\'c, F. M. Peeters, and K. L. Janssens,
Phys. Rev. B {\bf 65}, 165333 (2002); M. Tadi\'c and F. M. Peeters,
Phys. Rev. B {\bf 70}, 195302 (2004).
\bibitem{vurgaftman2001}I. Vurgaftman, J. R. Meyer, and L. R.
Ram-Mohan J. Appl. Phys. {\bf 89}, 5815 (2001).
\bibitem{davies1997}J. H. Davies, {\it The Physics of Low-dimensional
Semiconductors} (Cambridge University Press, 1997).
\bibitem{climente2003}J. I. Climente, J. Planelles, and W. Jask\'olski,
Phys. Rev. B {\bf 68}, 075307 (2003).
\bibitem{efros1996}Al. L. Efros, M. Rosen, M. Kuno, M. Nirmal, D. J. Norris,
and M. Bawendi, Phys. Rev. B {\bf 54}, 4843 (1996).
\bibitem{stier1999}O. Stier, M. Grundmann, and D. Bimberg, Phys. Rev. B
{\bf 59}, 5688 (1999).
\bibitem{bayer1999}M. Bayer, A. Kuther, A. Forchel, A. Gorbunov, V. B. Timofeev,
F. Schäfer, and J. P. Reithmaier, Phys. Rev. Lett. {\bf 82}, 1748 (1999).
\end{thebibliography}
\end{document}